\documentclass[journal,10pt]{IEEEtran}
\IEEEoverridecommandlockouts

\usepackage{array}
\usepackage{amsbsy}
\usepackage{amsthm}
\usepackage{amssymb}
\usepackage{amsmath}
\usepackage{multirow}
\usepackage{epsfig}
\usepackage{subfigure}
\usepackage{graphics}
\usepackage{multicol}
\usepackage{diagbox}
\usepackage{upgreek}
\usepackage{threeparttable}

\usepackage{epstopdf}

\usepackage{makecell}
\usepackage[]{pifont}
\usepackage{epstopdf}
\usepackage{balance}
\usepackage{xcolor}
\usepackage{cite}
\usepackage{color}
\usepackage{algorithm}
\usepackage{algorithmic}
\usepackage{textcomp}
\usepackage{subfigure}
\usepackage{subfloat}
\usepackage{booktabs}
\usepackage{lipsum}
\usepackage{bm}
\usepackage{filecontents}

\usepackage{graphicx}

\makeatletter
\newcommand*\bigcdot{\mathpalette\bigcdot@{.5}}
\newcommand*\bigcdot@[2]{\mathbin{\vcenter{\hbox{\scalebox{#2}{$\m@th#1\bullet$}}}}}
\makeatother

\usepackage{scalerel}
\usepackage{stackengine}
\usepackage{pgf}
\newcounter{iloop}
\newcommand\openbigstar[1][0.7]{%
  \scalerel*{%
    \stackinset{c}{-.125pt}{c}{}{\scalebox{#1}{\color{white}{$\bigstar$}}}{%
      $\bigstar$}%
  }{\bigstar}
}
\newcommand{\Stars}[1]{\ensuremath{%
\pgfmathtruncatemacro{\imax}{ifthenelse(int(#1)==#1,#1-1,#1)}%
\pgfmathsetmacro{\xrest}{0.9*(1-#1+\imax)}%
\setcounter{iloop}{0}%
\loop\stepcounter{iloop}\ifnum\value{iloop}<\the\numexpr\imax+1
\bigstar\repeat
\openbigstar[\xrest]%
\setcounter{iloop}{0}%
\loop\stepcounter{iloop}\ifnum\value{iloop}<\the\numexpr5-\imax\relax
\openbigstar[.9]\repeat}}

\usepackage{colortbl}
\usepackage{xcolor}

\graphicspath{ {figures/}}

\usepackage{hyperref}
\hypersetup{
    colorlinks=true,
    linkcolor=blue,
    filecolor=blue,      
    urlcolor=blue,
    citecolor=cyan,
}
\begin{document}

\title{{\huge Rethinking Signaling Design for ISAC: \\From Pilot-Based  to Payload-Based Sensing}}

 \author{Yunxin Li,~\IEEEmembership{Graduate Student Member,~IEEE}, Ying Zhang,~\IEEEmembership{Graduate Student Member,~IEEE}, \\ Christos Masouros,~\IEEEmembership{Fellow,~IEEE}, Sofie Pollin,~\IEEEmembership{Senior Member,~IEEE}, Fan Liu,~\IEEEmembership{Senior Member,~IEEE}

\vspace{-1.5em}

\IEEEcompsocitemizethanks{
\IEEEcompsocthanksitem Yunxin Li is with the School of Automation and Intelligent Manufacturing (AIM), Southern University of Science and Technology, Shenzhen 518055, China, and also with the Department of Electrical Engineering (ESAT), KU Leuven, 3000 Leuven, Belgium.\protect\\
E-mail: liyx2022@mail.sustech.edu.cn.
\IEEEcompsocthanksitem Ying Zhang is with the School of Automation and Intelligent Manufacturing (AIM), Southern University of Science and Technology, Shenzhen 518055, China.\protect\\
E-mail: zhangying2024@mail.sustech.edu.cn.
\IEEEcompsocthanksitem Christos Masouros is with the Department of Electronic and Electrical Engineering, University College London, London WC1E 7JE, U.K. \protect\\
E-mail: c.masouros@ucl.ac.uk.
\IEEEcompsocthanksitem Sofie Pollin is with the Department of Electrical Engineering, KU Leuven, 3000 Leuven, Belgium. \protect\\
E-mail: sofie.pollin@kuleuven.be.
\IEEEcompsocthanksitem Fan Liu is with the National Mobile Communications Research Laboratory, Southeast University, Nanjing 210096, China.\protect\\
E-mail: fan.liu@seu.edu.cn.
}% <-this % stops a space
\thanks{(Corresponding author: Fan Liu.)}

}

% make the title area
\maketitle

\begin{abstract}
Integrated Sensing and Communications (ISAC) is emerging as a key enabler for 6G networks, with signaling design at the core of its evolution. This paper reviews the paradigm shift of ISAC signaling designs from pilot-aided sensing to data payload-based approaches, with a particular focus on how these techniques can be realized within existing 5G NR structures. We commence with the reuse of pilots and reference signals that exploit existing 5G New Radio (NR) structures for sensing. Then, we extend to more advanced approaches that integrate the data payload through novel constellation shaping, modulation bases, and pulse shaping filters. We highlight the opportunities and tradeoffs that arise when extending sensing from sparse pilot and reference signal resources to the full communication frame, emphasizing how constellation properties and modulation choices directly determine sensing performance. To illustrate practical feasibility, a case study on sensing-assisted NR Vehicle-to-Infrastructure (V2I) networks demonstrates how exploiting both reference signals and payload echoes can reduce signaling overhead and enable proactive beam management and handover.
\end{abstract}

\IEEEpeerreviewmaketitle

\section{Introduction}\label{Introduction}

Integrated Sensing and Communications (ISAC) has rapidly emerged as a defining feature of the envisioned 6G landscape, gaining formal recognition by ITU-R within the IMT-2030 framework as a key usage scenario~\cite{itu2023framework}. Complementing this, standardization bodies such as ETSI have launched the ISAC Industry Specification Group (ISG) to develop ISAC use cases, evaluation methodologies, channel models, and architectural frameworks for 6G deployment. Moreover, 3GPP’s System Aspects Working Group 1 (SA1) and Radio Access Network (RAN) groups laid the groundwork for ISAC in Release 19 and beyond, examining how existing interfaces and functions can support sensing integration.

At the heart of ISAC lies signaling design. Broadly, there are two main approaches to achieving joint communication and sensing functions: \textit{Sensing-Centric Signaling Design} and \textit{Communication-Centric Signaling Design}. Sensing-centric signals are based on classic radar formats such as frequency-modulated continuous wave (FMCW) or chirp signals~\cite{mishra2019SPM}. These formats are highly effective for high-resolution sensing, but do not align directly with cellular communication standards like 5G NR, leading to low data throughput and poor compatibility. Incorporating communication into these signals via amplitude or phase modulation of chirps has been studied, but such schemes continue to suffer from limited data rates dictated by radar pulse repetition frequencies. Communication-centric signals, by contrast, embed sensing into existing communication structures such as OFDM frames. This approach maximizes standard compliance and reduces the complexity of implementation. Within this paradigm, sensing and communication are typically separated in time or frequency, for instance by allocating certain time frames or Physical Resource Blocks (PRBs) to sensing. Alternatively, they can be integrated by using a unified signaling design. A key challenge in such designs is the deterministic–random tradeoff (DRT): sensing prefers deterministic sequences because they create sharp, predictable correlation peaks for accurate estimation, whereas communication requires random data symbols to encode information reliably. While existing approaches attempt to balance these two requirements and preserve backward compatibility, they often face sensing performance constraints compared to dedicated sensing-centric designs~\cite{keskin2025fundamental}.

Alongside these, new communication modulation basis such as Orthogonal Time Frequency Space (OTFS), Orthogonal Delay Division Multiplexing (ODDM) and Affine Frequency Division Multiplexing (AFDM) have emerged in recent years~\cite{rou2024orthogonal}. These signals exploit structures like delay-Doppler representations, chirp-based subcarriers, and doubly dispersive channel resilience to deliver enhanced performance under high-mobility and complex channel conditions. Despite their demonstrated capabilities, these signals are not yet part of 3GPP standards, but they may serve as strong candidates for 6G and beyond.

\begin{figure*}[h!]
    \centering
    \includegraphics[width=\linewidth]{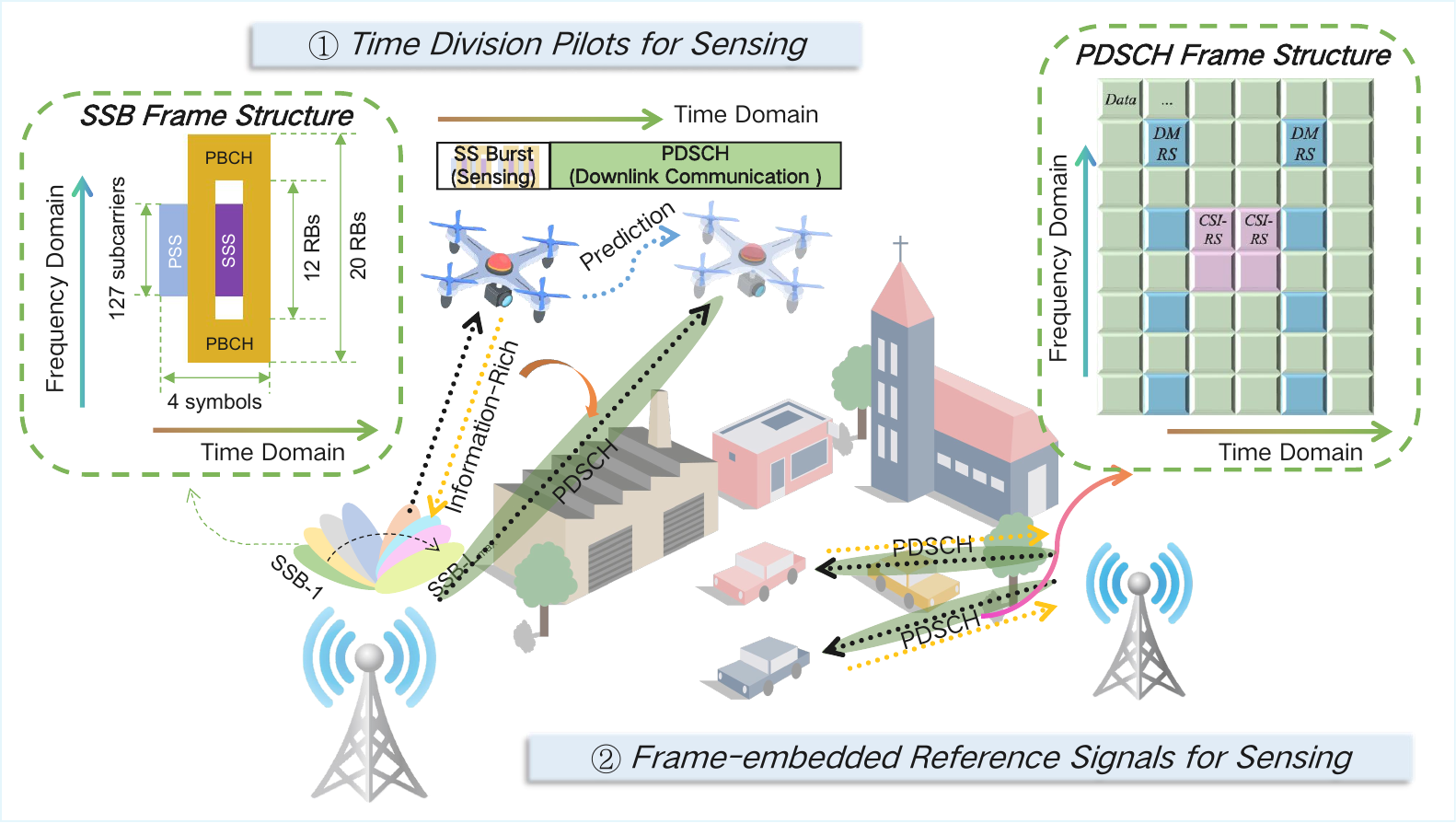}
    \caption{Sensing with pilots and reference signals.}
    \label{Pilots}
\end{figure*}

Building on this context, in this paper we overview signaling designs for ISAC spanning from pilot-based to data-payload extended designs. The remainder of this paper first reviews how sensing can be realized using existing 5G NR pilots and reference signals, highlighting both their strengths and inherent limitations. It then explores advanced payload-based sensing, emphasizing how constellation symbols, orthonormal modulation bases (e.g., OFDM, OTFS, AFDM), and pulse shaping filters influence the DRT between communication and sensing performance. Finally, a case study on sensing-assisted 5G NR V2I networks illustrates how ISAC, leveraging the standard OFDM signal along with embedded reference signals, can enhance four critical protocol stages (e.g., initial access, connected mode, beam failure and recovery, handover), offering pathways to reduce overhead and latency while maintaining compatibility with current standards.

\section{Sensing with Pilots and Reference Signals}

In current ISAC systems, sensing is primarily achieved using pilots or reference signals embedded in the communication frame. These signals, originally designed for channel estimation and synchronization, can be reused for sensing tasks such as range and velocity estimation. Their deterministic nature and well-defined sequence structures, such as those of Zadoff-Chu and Gold sequences, provide favorable autocorrelation properties, which are critical for accurate sensing.

In 5G NR, several standardized pilots and reference signals, including Synchronization Signal (SS) bursts, Demodulation Reference Signals (DMRS), Channel State Information Reference Signals (CSI-RS), and Sounding Reference Signals (SRS), are available for enabling sensing functionality~\cite{cui2022integrated,wei2024multiple}. As these signals are already embedded in standard 5G NR frames to support various communication tasks, they introduce minimal changes to the communication stack, thereby preserving compatibility with existing protocols. This reuse is particularly advantageous in early-stage ISAC deployments, where sensing capabilities can be integrated into existing 5G systems with limited architectural modifications.

\subsection{Sensing with Time-Division Pilots: SS Bursts}

One practical method for enabling sensing within existing communication standards is time division between sensing and communication, where specific time slots are reserved for sensing functionality. In 5G NR, a key component for this is the SS burst set, defined in 3GPP TS 38.211~\cite{3gpp.38.211}, which plays a critical role in initial access and beam management. Although the SS burst is primarily designed for synchronization purposes, its transmission occurs in dedicated symbols preceding the data payload, allowing sensing to be performed in these symbols with minimal interference from data transmission. In this sense, SS-based sensing can be regarded as a quasi time-division approach.

An SS burst set consists of a series of Synchronization Signal Blocks (SSBs), each containing a Primary Synchronization Signal (PSS), Secondary Synchronization Signal (SSS), and a Physical Broadcast Channel (PBCH). These blocks are transmitted across different spatial directions to facilitate directional cell discovery. The number of SSBs per burst set can be up to 64 in mmWave (FR2) deployments and up to 4 or 8 in sub-6 GHz (FR1), depending on deployment configurations. In the time domain, SS burst sets are transmitted periodically with 20 ms being a common default in commercial systems, and each burst set spans a time window of 5 ms. In the frequency domain, each SSB occupies 240 subcarriers and spans 4 OFDM symbols. 

Due to their fixed periodicity and wide spatial coverage via beam sweeping, SS bursts have recently been investigated for their potential in coarse sensing applications, such as environmental awareness or target detection. However, the relatively low repetition rate and short duration of SS bursts limit their temporal resolution. As a result, they are typically suitable for detecting large-scale or slowly varying environmental changes rather than fine-grained or high-speed motion. Even in such cases, improving the sensing performance typically requires extra frames and added latency.

\subsection{Sensing with Frame-Embedded Reference Signals: DMRS, CSI-RS, and SRS}

Another approach to enable sensing during ongoing communication is to reuse standardized reference signals embedded in the 5G NR frame. These signals, defined in 3GPP TS 38.211~\cite{3gpp.38.211}, occupy known time-frequency positions and use deterministic sequences, allowing the extraction of channel or environmental information without altering the signal.

\begin{itemize}
\item \textbf{DMRS}: DMRS, primarily used for demodulation and constructed using Zadoff-Chu or Gold sequences, is transmitted alongside data in both downlink (PDSCH) and uplink (PUSCH), typically occupying one or two OFDM symbols per slot, with optional additional DMRS symbols in high‐mobility scenarios. In the frequency domain, DMRS spans the entire allocated bandwidth portion in a comb pattern. This wide and dense frequency coverage and regular temporal presence make DMRS a strong candidate for sensing, enabling fine-grained delay and Doppler estimation as demonstrated in several ISAC studies.

\item \textbf{CSI-RS and SRS}: CSI‑RS is a configurable downlink reference signal used for channel state measurement and beam management. Its placement in time and frequency is defined via Radio Resource Control (RRC) parameters and varies in density, ranging from sparse to moderate, within selected PRBs and OFDM symbols. SRS, the uplink counterpart, is transmitted by the UE to sound the channel across frequency bands using comb-type subcarrier patterns and configurable periodicity. Both signals use orthogonal-coded sequences across antenna ports and are deployed less frequently than DMRS. Consequently, while they support spatial channel estimation and mobile measurements, their sparse time–frequency presence limits their suitability for continuous sensing.

\end{itemize}

\subsection{Limitations and Tradeoffs}

Pilots and reference signals are distributed across the time-frequency plane and are known to both the transmitter and receiver, making them well suited for sensing tasks during active communication, enabling environmental awareness without pausing data transmission. However, relying solely on these signals introduces several limitations, especially in high-mobility environments or scenarios requiring fine resolution.

\begin{itemize}
\item \textbf{Partial Resource Usage}:
These pilots and reference signals are intentionally designed to minimize overhead. For example, an SS burst occupies only a 5~ms window every 20~ms, DMRS typically spans 1–2 OFDM symbols per slot, and CSI‑RS is configured over a subset of subcarriers or periodic time slots. In general, these signals typically consume 5–15\% of the overall time-frequency resources. With so few observations available for coherent integration, the sensing process suffers from low effective SNR, making it difficult to detect weak targets.

\item \textbf{Sparse Time-Frequency Occupation}:
The sparse placement of reference signals across the time–frequency grid results in limited delay–Doppler resolution, restricting the ability to distinguish closely spaced objects or track fast motion. Recent work such as~\cite{keskin2021limited} shows that optimized sparse resource allocation can achieve strong estimation performance, but such adaptive designs are not supported in current NR configurations. Increasing the density could improve sensing resolution, but this comes at the cost of reduced spectral efficiency and increased signaling overhead, which is an undesirable tradeoff for most communication systems.

\end{itemize}

To overcome these limitations, recent ISAC research has begun exploring methods for performing sensing over the full communication frame, including both the pilots and payload data. While this introduces challenges, such as dealing with random data symbols and the need for advanced signal processing, it also opens up new opportunities for higher sensing resolution, improved environmental awareness, and more efficient protocol design. This transition from pilot-centric to payload-centric sensing becomes even more critical as future 6G systems may reduce the density of pilots to reduce overhead, thereby placing greater reliance on payload-centric designs to sustain high-resolution sensing. In the next section, we discuss how emerging signaling designs can support this transition.

\section{Sensing With Random Payload Signals}

By capitalizing on the complete spectral and temporal resources of the communication data payload, this approach primarily enhances sensing SNR and observation continuity, which in turn improve Doppler resolution, detection reliability, and parameter estimation fidelity. Although employing data payload signals for sensing can significantly outperform traditional pilot-based schemes, these signals are not inherently designed for sensing tasks, presenting critical challenges for communication-centric ISAC systems. Unlike the deterministic nature of conventional radar signals, data payload signals exhibit intrinsic randomness. This randomness enhances communication rates but degrades sensing performance, leading to the DRT between sensing and communication \cite{10471902}. Developed primarily for data transmission, communication signals differ fundamentally in structure from traditional radar signals. In practice, a communication signal can be decomposed into several essential components after channel coding:

\begin{figure*}[ht!]
    \centering
    \includegraphics[width=\linewidth]{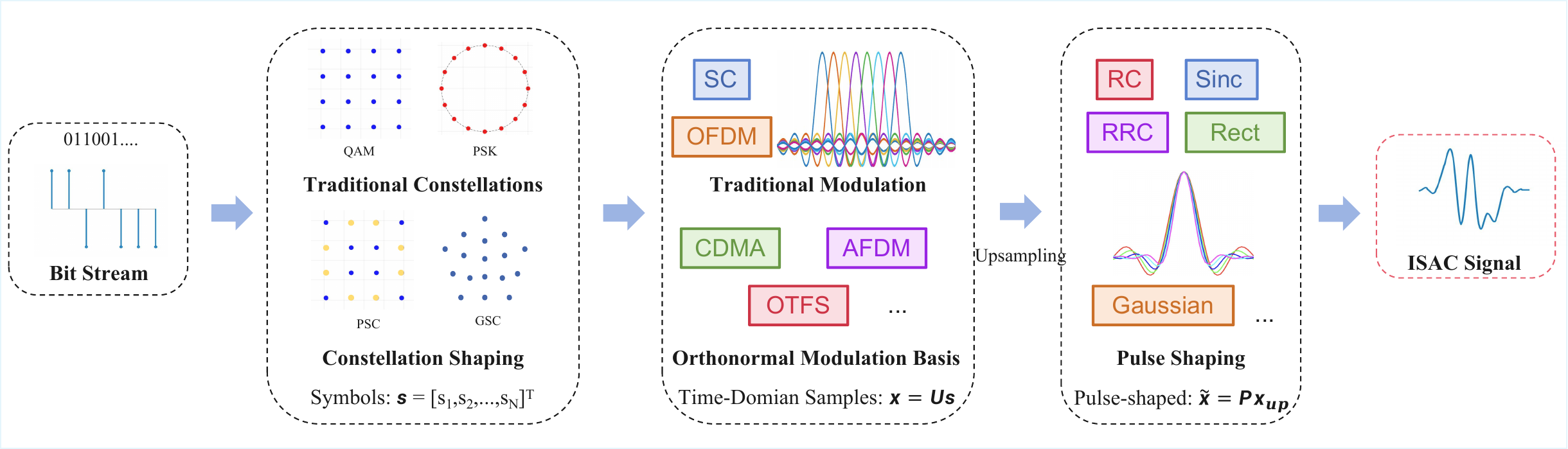}
    \caption{A signal processing pipeline for ISAC signal generation.}
    \label{Process}
\end{figure*}

\begin{itemize}
\item \textbf{Constellation Symbols} : Constellation symbols obtained through mapping bit sequences, represented by specific amplitude and phase values, which encapsulate the transmitted information.

\item \textbf{Orthonormal Modulation Basis} : An orthonormal set of modulation basis functions that transmit these symbols by constructing discrete-time signals.

\item \textbf{Pulse Shaping Filter} : Pulse shaping is applied through a filter that converts discrete samples in the time domain into their continuous counterparts.
\end{itemize}

Each of these components plays a significant role in influencing the achievable sensing performance.

\subsection{Constellation Symbols}
Constellation symbols, represented as discrete complex points, serve as fundamental units carrying information. Each symbol maps to a unique bit pattern, and the geometry of the constellation significantly affects spectral efficiency, error performance, and system reliability. Common constellation types include Phase Shift Keying (PSK), Quadrature Amplitude Modulation (QAM), and Amplitude Phase Shift Keying (APSK). PSK conveys information via phase shifts, with variants like BPSK and QPSK valued for their robustness in interference-limited scenarios. QAM combines amplitude and phase modulation to enhance spectral efficiency, exemplified by formats such as 16QAM and 64QAM. APSK arranges constellation points on multiple concentric rings, offering a trade-off between power efficiency and spectral efficiency, making it particularly advantageous in satellite communication systems.

Recent studies revealed that the fourth-order moment (\textit{kurtosis}) of constellation symbols significantly affects the variance of the ambiguity function (AF) of communication signals. Intuitively, the AF is sensitive to symbol-to-symbol fluctuations. Constellations with high kurtosis produce occasional large-amplitude symbols, making the AF vary much more. Constellations with low kurtosis keep all symbols similar in magnitude, giving a smoother, more stable AF, which benefits sensing performance~\cite{11087656}. The standard complex Gaussian distribution, characterized by a kurtosis of 2, serves as a theoretical benchmark in ISAC systems. Based on this benchmark, constellations are classified into sub-Gaussian (kurtosis less than 2) and super-Gaussian (kurtosis greater than 2) categories. Notably, QAM, PSK and standard APSK all fall within the sub-Gaussian class, with QAM and standard APSK constellations exhibiting kurtosis values between 1 and 2, and PSK having kurtosis exactly equal to one, making them better suited for sensing.

Conventional constellation designs primarily target communication performance optimization, which may be insufficient to meet the stringent sensing accuracy demands of ISAC systems. Since ISAC requires simultaneous reliable data transmission and accurate sensing, communication-centric constellation designs inherently limit sensing capabilities. By employing Probabilistic Constellation Shaping (PCS) and Geometric Constellation Shaping (GCS) techniques\cite{10685511}—differing fundamentally in implementation—joint optimization of communication and sensing performance is achieved through adjusting the probability distribution and geometric arrangement of constellation points, respectively. 

\begin{itemize}
    \item $\textbf{PCS:}$ PCS achieves performance enhancement by adjusting the transmission probabilities of constellation points without altering their geometric configuration. This approach enables the system to approach channel capacity with relatively low implementation complexity and minimal modifications to existing hardware architectures and modulation detectors.

    \item $\textbf{GCS:}$ GCS involves redesigning the spatial arrangement of constellation points within the complex plane, thereby modifying the constellation geometry to optimize key metrics such as the minimum Euclidean distance and power distribution. Although this method can significantly improve error performance, it typically entails increased signal processing complexity and more demanding system design requirements.
\end{itemize}
Both methods can be applied independently or combined to enhance communication and sensing performance, or trade off between them, based on system requirements.

\subsection{Orthonormal Modulation Basis}
In conventional communication systems, a sequence of symbols is modulated over an orthonormal basis defined in the time domain. For clarity, several representative examples are enumerated as follows:

\begin{itemize}
\item $\textbf{Single Carrier (SC):}$ SC transmits symbols sequentially with low peak-to-average power ratio (PAPR), offering robustness to phase noise and efficient power usage. Its main limitation lies in inter-symbol interference (ISI) under frequency-selective fading, which affects both communication and sensing accuracy.  Despite these challenges, SC has been widely adopted in systems such as IEEE 802.11ad/ay WLAN, where hardware simplicity and energy efficiency are important. In cellular networks, transmissions have shifted toward OFDM, but SC remains attractive in contexts demanding low complexity and high power efficiency.
\item $\textbf{OFDM:}$ OFDM modulates symbols across multiple frequency subcarriers using an inverse discrete Fourier transform (IDFT) basis, achieving high spectral efficiency and robustness to multipath environments. Due to its flexibility and reliability, OFDM is widely used in modern communication systems like LTE, Wi-Fi, and 5G. Numerous studies have investigated the sensing performance and associated algorithms of OFDM, and its superior sensing capabilities have been rigorously demonstrated.
\item $\textbf{Code-Division Multiple Access (CDMA):}$ CDMA spreads symbols with pseudo-random codes, enabling multiuser access and interference resilience. The autocorrelation and cross-correlation properties of these codes are also useful for sensing, as they allow reliable detection and separation of multiple targets.
\item $\textbf{OTFS:}$ OTFS maps symbols in the delay–Doppler domain via inverse symplectic finite Fourier transform (ISFFT), spreading each symbol across the time–frequency plane. This makes it highly robust to Doppler shifts and fast-varying channels, yielding improved sensing resolution and stable delay–Doppler parameter estimation in high-mobility scenarios.
\item $\textbf{AFDM:}$ AFDM employs orthogonal chirp signals in the affine Fourier transform (AFT) domain, naturally exploiting the channel’s delay–Doppler structure. An AFDM-enabled ISAC framework with a superimposed pilot design was introduced in~\cite{11164334}, analytically shown to enhance ambiguity-function sharpness and channel estimation, making AFDM a promising candidate for ISAC in 6G network.
\end{itemize}

In principle, any communication signal can be adapted for sensing purposes; however, this reuse is subject to inherent limitations. Since these signals are primarily optimized for communication performance, the added sensing functionality often experiences degraded performance. Despite these drawbacks, the communication-centric approach remains appealing owing to its low complexity, standards compatibility, and practical feasibility for real-world implementation. While emerging modulation bases such as OTFS and AFDM offer strong potential for ISAC in 6G, their integration with current NR systems remains non-trivial. Because these waveforms operate in the delay–Doppler or affine-frequency domains rather than the time–frequency grid of OFDM, they would require new resource mapping, synchronization, and channel estimation procedures, as well as modifications to the baseband processing chain. These integration challenges represent an important step toward backward compatible adoption within standardized frameworks.

\begin{figure}[ht!]
    \centering    
    \subfigure[The P-ACF of the 16-QAM constellation under various signaling schemes (1024 symbols).]{\includegraphics[width=0.9\linewidth]{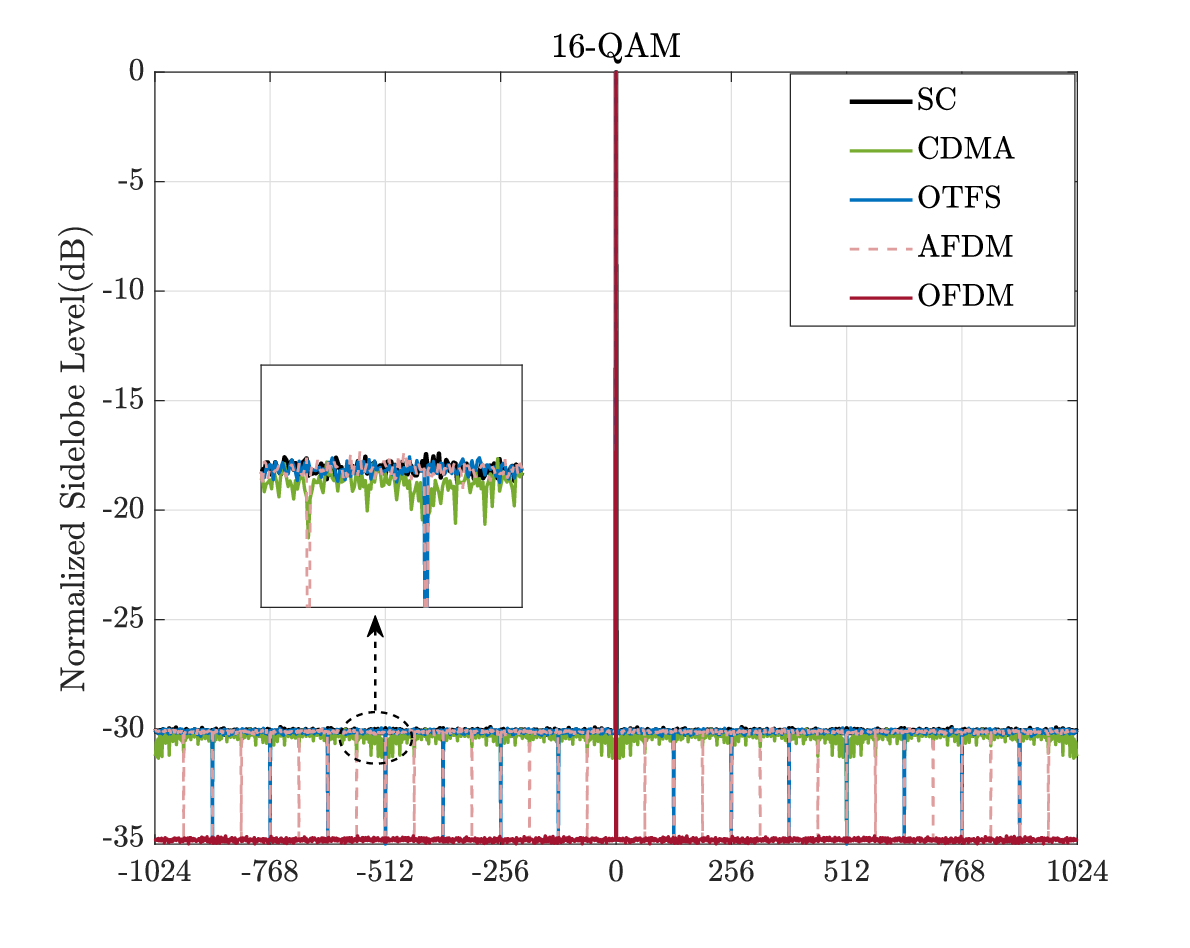}
    \label{16QAM}}
    \subfigure[The average squared P-ACFs with no coherent integration and 10,000 coherent integrations under OFDM signaling with 16-QAM constellation and RRC filter.]{\includegraphics[width=0.9\linewidth]{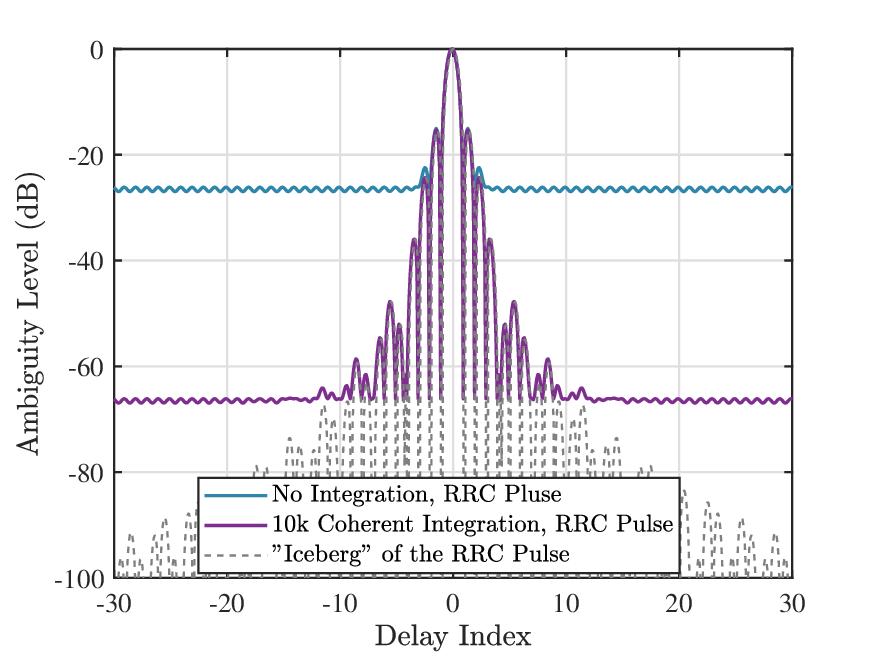}
    \label{RRC}}
    \caption{The P-ACF results with different modulation basis and the P-ACF results with RRC pulse filter.}
\end{figure}

Recent theoretical advances have provided deeper insight into this tradeoff, especially regarding the design of modulation bases and their impact on sensing performance. As illustrated in Fig.~\ref{16QAM}, the periodic auto correlation function (P-ACF) of the 16-QAM constellation under different signaling schemes reveals that sidelobe characteristics, which directly determine ranging accuracy, vary significantly with the modulation basis. Among different schemes, OFDM has been shown to achieve the lowest possible average ranging sidelobe levels under sub-Gaussian constellations such as QAM, and its sidelobes vanish entirely under PSK modulation~\cite{11087656}. These results establish OFDM not only as a practical but also a theoretically optimal choice for minimizing sidelobe levels, thereby ensuring reliable delay estimation in ISAC systems. Such findings confirm that the interplay between constellation properties and modulation bases is more than an implementation detail: it defines the achievable limits of joint communication and sensing, and provides a rigorous foundation for signaling design in 6G and beyond.

\subsection{Pulse Shaping Filter}
In communication systems, pulse shaping is a key baseband signal processing technique designed to efficiently transmit data within limited bandwidth while reducing inter-symbol interference (ISI). In digital modulation, information bits are mapped to constellation symbols and converted into discrete-time pulses. Directly transmitting rectangular pulses leads to severe spectral leakage and ISI at the receiver, degrading system performance. Pulse shaping applies filtering at the transmitter to confine the signal’s bandwidth in the frequency domain and meet the Nyquist criterion in the time domain, eliminating ISI at ideal sampling points.  Pulse shaping not only affects spectral efficiency and interference immunity but also impacts power efficiency and receiver complexity, making it indispensable in the design of high-speed, broadband, and spectrum-constrained communication systems.

Widely used pulse-shaping filters include the Sinc filter, Gaussian filter, Raised Cosine (RC) filter, and Root Raised Cosine (RRC) filter. The Sinc filter is the ideal Nyquist pulse shaping filter, featuring a rectangular frequency response and an infinite-length sinc function in the time domain. The Sinc pulse constitutes the most fundamental pulse-shaping filter. However, its requirement for an ideal rectangular spectrum renders it practically infeasible. The Gaussian pulse filter shapes signals using a Gaussian function, providing smooth spectral characteristics. Yet its broad main lobe and nonzero sidelobes limit spectral efficiency and may cause inter-symbol interference. The RC and RRC filters are widely used for pulse shaping to reduce inter-symbol interference. Nevertheless, both filters have relatively long impulse responses, which can increase implementation complexity.

In the context of ISAC, pulse shaping also plays a critical role in sensing performance, as it determines the statistical properties of the auto-correlation function (ACF) of random ISAC signals~\cite{11037613}. As analyzed, the expectation of the squared ACF can be metaphorically described as “an iceberg in the sea”, where the “iceberg” represents the squared mean of the ACF, shaped by the squared ACF of the pulse shaping filter itself, defining the mainlobe and near sidelobes. The “sea level” characterizes the variance arising from data payload randomness, affecting distant sidelobes. The “sea level” can be reduced through coherent integration, but the ultimate sensing performance, including ranging resolution and sidelobe suppression, depends on the pulse design. By 'iceberg shaping', in other words, optimizing Nyquist pulses to reduce sidelobe levels in specified delay regions, the performance of target detection and multi-target resolution can be improved.

As shown in Fig.~\ref{RRC}, the average squared ACF of a pulse-shaped OFDM signal with the squared ACF of the pulse itself are compared, where the constellation used is 16-QAM and the pulse shaping filter is the RRC filter. The results show that the ACF of the random OFDM signal closely matches the ``tip'' of the iceberg, which is the ACF of the pulse. Beyond the ``tip'' of the iceberg, the sidelobe level is primarily attributed to the “sea level” component. Further, the ACF obtained from $10000$ coherent integrations is compared with its no-integration counterpart, with all other parameters remaining consistent. It's obvious that the ``sea level'' is reduced and more of the ``iceberg'' is unveiled. It means that the sidelobe level of the ACF can be reduced by coherent integration, thereby improving the sensing performance, ultimately limited by the sidelobes due to the PSF.

Furthermore, since both channel estimation and sensing parameter extraction rely on similar correlation-based operations, combining deterministic reference signals with random payload symbols in OFDM-based NR V2I systems may further enhance sensing accuracy. Pilots provide strong autocorrelation for reliable delay–Doppler estimation, while payload symbols contribute additional energy that improves detection sensitivity. Potential advanced strategies include weighted joint correlation of pilots and payloads, and learning-based fusion methods that exploit the complementary structure of the two components to achieve superior sensing performance without modifying the NR frame design.

\section{Case study: ISAC-enabled OFDM NR V2I Networks}

\begin{figure*}[htbp]
    \centering
    \includegraphics[width=\linewidth]{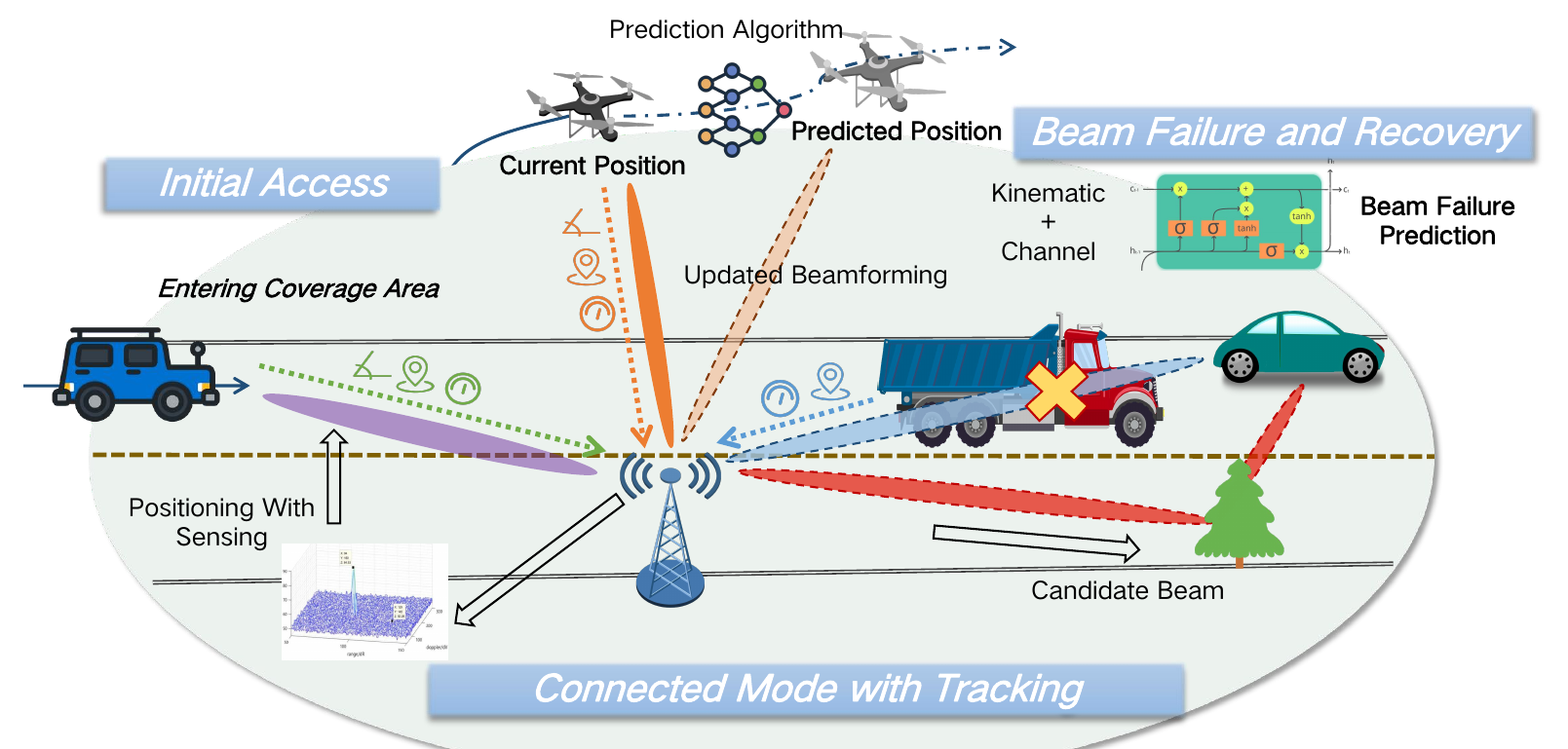}
    \caption{Sensing-assisted beam management.}
    \label{Beam management}
\end{figure*}

As the dominant communication signal in current mobile standards, OFDM is expected to continue playing a foundational role in 6G systems for its backward compatibility with 5G NR and its proven optimality for ranging tasks with QAM/PSK constellations as discussed in the previous section. At the same time, the increasing demands of high-mobility environments, such as vehicle-to-infrastructure (V2I) communications, expose the limitations of conventional communication-centric protocols. These include frequent beam training, delayed reaction to link degradation, and signaling-heavy handover processes. 

In this case study, we concretely implement the previously discussed ISAC signaling concepts by jointly exploiting NR reference signals and payload symbols within the standard OFDM frame, thereby validating the achievability of hybrid sensing under existing 3GPP NR structures. Reference signals, embedded within the payload grid, provide deterministic sequences that enable accurate channel estimation and data demodulation, while the surrounding payload data contribute additionally that span the full time–frequency resources. By jointly leveraging these two components of the same OFDM block and the echo of the entire frame, the gNB can extract angle, delay, and Doppler information with finer resolution and higher continuity. This reuse of the whole frame reduces dependence on signaling-heavy procedures, lowers overhead, and preserves full compliance with the NR frame structure. The following subsections illustrate how this approach benefits four critical stages in NR V2I networks.

\subsection{Initial Access}
In conventional NR systems, initial access relies on broadcasting multiple SSBs to sweep the entire spatial domain, with up to 64 beams transmitted every 20~ms in mmWave deployments. This exhaustive beam sweeping introduces latency and consumes valuable resources, especially at higher frequencies.

With sensing capabilities, the gNB can estimate the presence, position, and motion of vehicular UEs by extracting angle, delay, and Doppler information from the echoes of the transmitted signal. Instead of sending a wide set of beams blindly, the gNB directs synchronization signals toward the estimated direction of the UE, greatly reducing sweeping overhead and latency from 15~ms to 1.25~ms, achieving up to 91.7\% reduction for 120~kHz subcarrier spacing~\cite{li2024frame}. This reduces the need for repeated sweeping, shortens access time, and increases the likelihood of a strong initial link, all while operating within the standard NR OFDM frame.

\subsection{Connected Mode}

Maintaining a stable connection with a moving vehicle requires frequent updates on the optimal beam direction. Traditionally, this is implicitly done by using channel state reference signals (e.g., CSI-RS), along with the associated periodic uplink feedback from the UE. These reference signals occupy bandwidth and time resources, thus reducing the data throughput and decreasing the spectrum efficiency. 

\begin{figure}[ht!]
    \centering    
    \includegraphics[width=0.9\linewidth]{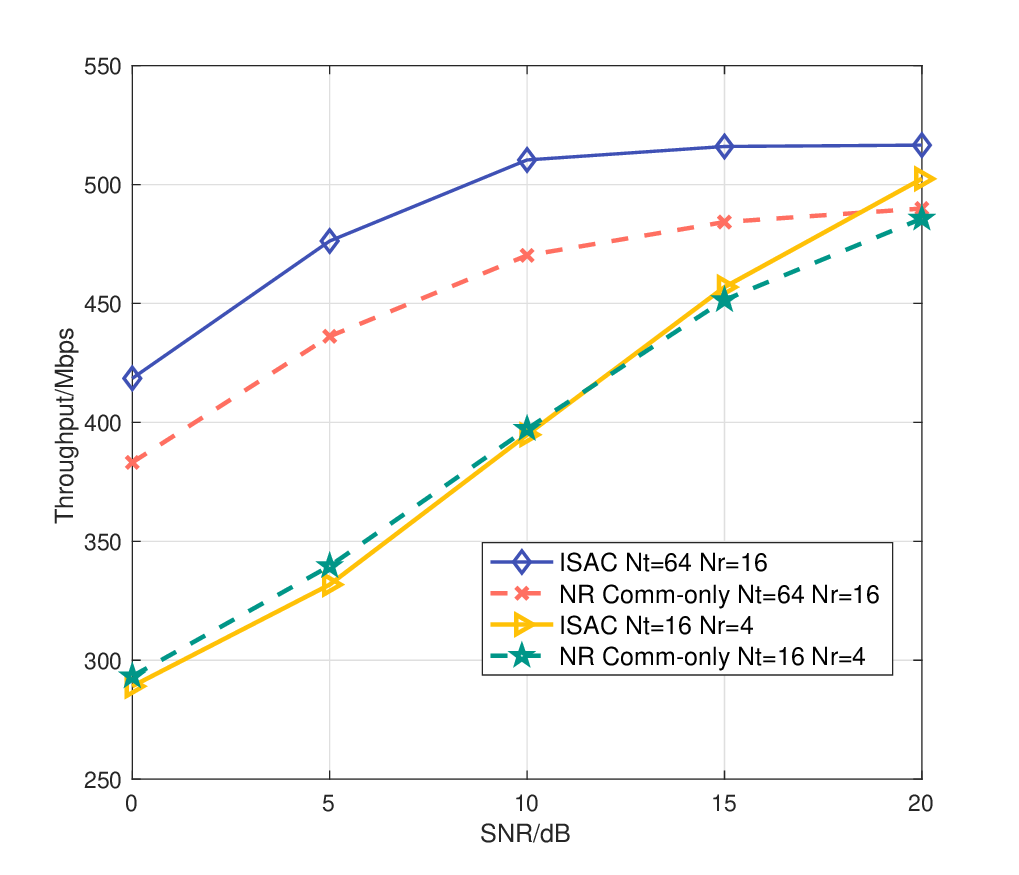}
    \caption{Throughput comparison in connected mode under sensing-assisted scheme and NR communication-only scheme.}
    \label{TP}
\end{figure}

By contrast, ISAC-enabled systems allow the gNB to monitor the vehicle’s motion and orientation extracted from the payload echoes. Techniques like Doppler and delay estimation can be applied to data-bearing OFDM blocks, enabling real-time tracking without depending heavily on pilot signals. As shown in Fig.~\ref{TP}, by leveraging accurate tracking algorithms, a portion of the reference signals and its uplink feedback can be omitted (e.g., CSI-RS in single layer SU-MIMO), reducing pilot signaling overhead and increasing throughput while still maintaining precise beam alignment, particularly important in environments where vehicles move rapidly and unpredictably.

\subsection{Beam Failure and Recovery}
Beam failures, caused by sudden blockages or rapid mobility, are traditionally detected via reference signal thresholds such as Reference Signal Received Power (RSRP) and Reference Signal Received Quality (RSRQ), triggering timers and the corresponding reactive recovery mechanisms. 

With the deployment of ISAC, the gNB can anticipate beam failures by continuously monitoring vehicle motion using information-rich echoes. Abrupt changes in velocity or range are detected earlier through sensing, enabling predictive recovery before the communication link fully degrades, outperforming conventional beam failure detection based on RSRP measurements, and reducing detection latency by approximately 50\%, from 5~ms to 2.5~ms for 120~kHz subcarrier spacing. By preparing candidate beams in advance using channel reciprocity, the network reduces outage time and ensures more reliable connectivity. This proactive strategy is especially valuable in V2I scenarios, where safety critical applications demand ultra low latency and uninterrupted service.

\begin{figure}[ht!]
    \centering    
    \includegraphics[width=\linewidth]{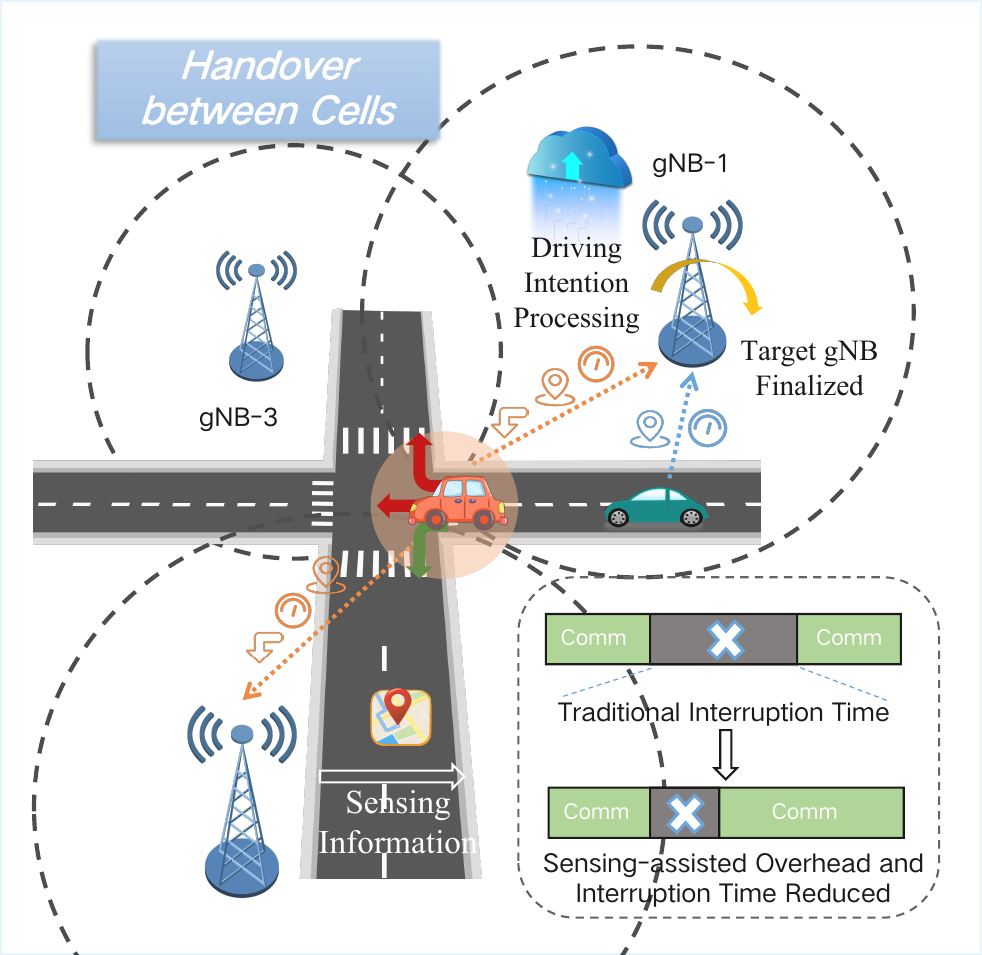}
    \caption{Sensing-assisted handover.}
    \label{Handover}
\end{figure}

\subsection{Handover between Cells}
Handover is a critical process in communication networks where users move across cell boundaries. In standard 5G NR systems, handover is typically triggered based on comparisons of received signal power from neighboring cells, combined with a time-to-trigger parameter. This reactive mechanism may cause late or unnecessary handovers, particularly in dynamic urban environments.

With sensing embedded in the communication signal, handover decisions can be informed by richer context. For example, changes in kinematic parameters extracted from the reflections of downlink signals allow the network to predict whether a vehicle at an intersection will turn or continue straight. Based on these predictions, the network can select the most suitable target cell and initiate preparations before the actual handover event. This predictive handover reduces interruption time and ensures more seamless connectivity, marking a step toward proactive mobility management in 6G networks.

\section{Research Directions}
Looking ahead, the integration of innovative ISAC signaling design into future standards will require balancing backward compatibility with new performance gains and will need more research on several critical advancements:
\begin{itemize}
    \item $\textbf{Unified Performance Metrics:}$ Currently, a key research gap is the lack of standardized performance measures that jointly capture sensing and communication efficacy. Bridging this gap via information-theoretic modeling and developing unified benchmarking frameworks will provide clarity in algorithm and signaling design.
    \item $\textbf{Integration with Coding Design:}$ By co-designing channel coding and ISAC signals, system can balance communication reliability and sensing accuracy, enabling improved tradeoffs between error protection and parameter estimation performance.
    \item $\textbf{Joint Signaling and Receiver Design:}$ Traditional designs treat signal generation and receiver processing separately. Co-designing signaling structures and advanced receivers that can extract both data and environmental parameters simultaneously while avoiding mutual interference will be key to achieving full ISAC gains.
    \item $\textbf{AI Adaptive Signals:}$ As networks grow more dynamic and complex, machine learning offers a compelling path to real-time adaptation of constellation shapes, modulation bases, and pulse filters. Such AI-driven optimization can enable flexibility under diverse channel conditions and heterogeneous deployment scenarios.
    \item $\textbf{Security and Privacy in ISAC:}$ The dual use nature in ISAC signal poses unique security and privacy concerns. Secure signaling design is essential to address these risks.
    \item $\textbf{Use Cases and Performance for Standardization:}$ ETSI’s latest ISAC report introduces advanced use cases (e.g., human motion detection, emergency response, autonomous vehicles), and proposes structured integration levels and sensing modes for standard alignment~\cite{ETSI}. These insights should guide signal-level requirements and deployment strategies.
\end{itemize}

\section{Conclusion}
This paper has examined the central role of signaling design in enabling ISAC for 6G. We showed that 5G NR pilots and reference signals provide an immediate path to sensing with minimal changes to existing systems. Beyond this, advanced signaling strategies that exploit payload data, constellation shaping, pulse filters, and novel modulation bases such as OTFS, CDMA, and AFDM can significantly improve sensing resolution and robustness. A case study on OFDM-based NR V2I networks further demonstrated how sensing can enhance beam management and handover while reducing overhead. Together, these discussions emphasize that signaling design will remain a cornerstone of ISAC research and standardization, shaping the foundation for 6G and beyond.

\bibliographystyle{IEEEtran}
\bibliography{ref}

\end{document}